\documentclass[11pt]{article}

\usepackage{epsfig}
\usepackage{latexsym}
\usepackage{enumerate}
\usepackage{url}
\usepackage{float}
\usepackage[hypertexnames=false,hyperfootnotes=false]{hyperref}
\usepackage{texnansi}
\usepackage{color}
\usepackage{tikz}
\usepackage[margin=10pt,font=small,labelfont=bf]{caption}
\usepackage[subrefformat=parens,labelformat=parens]{subcaption}
\usepackage{afterpage}
\usepackage{enumitem}
\usepackage[boxed]{algorithm}
\usepackage{algpseudocode}
\usepackage[normalem]{ulem}
\usepackage{lmodern}
\usepackage{booktabs}
\usepackage{sectsty}
\usepackage{ifthen}
\usepackage{amsmath}
\usepackage{amsthm}
\usepackage{amssymb}
\usepackage{amsfonts}
\usepackage{thmtools}
\usepackage{thm-restate}
\usepackage{mathtools}
\usepackage{xspace}
\usepackage{titling}
\usepackage{natbib}
\usepackage{xfrac}
\usepackage{multirow}
\usepackage{bigdelim}
\usepackage{bm}
\usepackage{cleveref}
\usepackage[textsize=tiny]{todonotes}
\newcommand{\I}[1]{\ensuremath{\mathbb{I}_{\left\{#1\right\}}}} % indicator function
\newcommand{\tends}{\ensuremath{\rightarrow}} % arrow for limits
\newcommand{\E}{\ensuremath{\mathsf{E}}} % expectation
\newcommand{\defeq}{\ensuremath{\triangleq}}
\DeclarePairedDelimiter\parens{\lparen}{\rparen}

\DeclarePairedDelimiter\bracks{\lbrack}{\rbrack}

\DeclareMathOperator*{\argmax}{\mathrm{argmax}}

\declaretheoremstyle[headfont=\sffamily\bfseries,bodyfont=\itshape]{thm-sf}
\declaretheorem[style=thm-sf]{theorem}

\declaretheorem[style=thm-sf]{assumption}
\crefname{assumption}{assumption}{assumptions}

\declaretheorem[style=thm-sf]{example}

\declaretheorem[style=thm-sf]{lemma}

\renewcommand{\thmcontinues}[1]{\hyperref[#1]{continued}}
\newcommand{\paraheader}[1]{\smallskip\noindent{\sffamily\bfseries #1}}
\usetikzlibrary{arrows,patterns,plotmarks,pgfplots.groupplots}
\tikzstyle{every picture} += [>=stealth]
\tikzset{axis/.style={semithick, line join=miter}}
\allsectionsfont{\sffamily}
\makeatletter
\def\@seccntformat#1{\csname the#1\endcsname.\quad}
\makeatother
\floatname{algorithm}{\normalfont\sffamily\bfseries Algorithm}
\floatstyle{ruled}
\newcommand{\emailhref}[1]{\href{mailto:#1}{\tt #1}} % hyperlinked email address
\provideboolean{fastcompile}
\newcommand{\hidefastcompile}[1]{\ifthenelse{\boolean{fastcompile}}{}{#1}}
\usepackage{pgfplots}
\usepackage{pgfplotstable}
\usetikzlibrary{calc}
\usepackage{mathtools}
\definecolor{orange}{rgb}{0.85,0.33,0.13} % 217,85,33
\definecolor{green}{rgb}{0.13,0.85,0.33}
\definecolor{purple}{rgb}{0.33,0.13,0.85}
\definecolor{lime}{rgb}{0.65,0.85,0.13}
\definecolor{blue}{rgb}{0.13,0.65,0.85}
\pgfplotscreateplotcyclelist{tricolor}{%
  orange,every mark/.append style={fill=orange!80!black},mark=*\\%
  green,every mark/.append style={fill=green!80!black},mark=square*\\%
  purple,every mark/.append style={fill=purple!80!black},mark=otimes*\\%
  black,mark=star\\%
  orange,every mark/.append style={fill=orange!80!black},mark=diamond*\\%
  green,densely dashed,every mark/.append style={solid,fill=green!80!black},mark=*\\%
  purple,densely dashed,every mark/.append style={solid,fill=purple!80!black},mark=square*\\%
  black,densely dashed,every mark/.append style={solid,fill=gray},mark=otimes*\\%
  orange,densely dashed,mark=star,every mark/.append style=solid\\%
  green,densely dashed,every mark/.append style={solid,fill=green!80!black},mark=diamond*\\%
}
\pgfplotsset{colormap={tricolormap}{color=(orange) color=(green) color=(purple)},
  colormap={quadcolormap}{color=(orange) color=(lime) color=(blue) color=(purple)}}
\pgfplotstableset{%
  font=\small,
  every head row/.style={before row=\toprule[1pt], after row=\midrule},
  every last row/.style={after row=\bottomrule[1pt]}}
\pgfplotsset{compat=1.15}
\usepackage{setspace}
\usepackage[margin=1in]{geometry}

\provideboolean{submissionversion}
\setboolean{submissionversion}{true}
\provideboolean{blindversion}
\ifthenelse{\boolean{submissionversion}}{
  \renewcommand{\todo}[2][1]{}
  
  \newcommand{\deledit}[1]{}
}{
  
  \newcommand{\deledit}[1]{{\color{orange} \sout{#1}}}
}

\newcommand{\pnl}{\ensuremath{\mathsf{P\&L}}\xspace}
\newcommand{\LVR}{\ensuremath{\mathsf{LVR}}\xspace}

\newcommand{\Q}{\ensuremath{\mathbb{Q}}\xspace}

\newcommand{\RENT}{\ensuremath{\mathsf{RENT}}\xspace}

\newcommand{\ARBPROFIT}{\ensuremath{\mathsf{ARB\_PROFIT}}\xspace}
\newcommand{\ARBEXCESS}{\ensuremath{\mathsf{ARB\_EXCESS}}\xspace}

\newcommand{\LFF}{\ensuremath{L_{\mathsf{ff}}}\xspace}

\newcommand{\PNLFF}{\ensuremath{\Pi^{\mathsf{LP}}_{\mathsf{ff}}}\xspace}
\newcommand{\PNLMGR}{\ensuremath{\Pi^{\mathsf{MGR}}_{\mathsf{am}}}\xspace}
\newcommand{\PNLLP}{\ensuremath{\Pi^{\mathsf{LP}}_{\mathsf{am}}}\xspace}

\newcommand{\AP}{\ensuremath{\mathsf{AP}}\xspace}
\renewcommand{\AE}{\ensuremath{\mathsf{AE}}\xspace}
\newcommand{\fopt}{\ensuremath{f_{\mathsf{opt}}}}

\ifthenelse{\boolean{blindversion}}{
  \title{\bf\sffamily am-AMM: An Auction-Managed Automated Market Maker}
  \author{\small Austin Adams \and \small Ciamac Moallemi \and \small Sara Reynolds \and \small Dan Robinson}
  \date{}
}{
  \title{\bf\sffamily am-AMM: An Auction-Managed Automated Market Maker\thanks{
      The second author is supported by the Briger Family Digital Finance Lab at Columbia
      Business School, and is an advisor to Paradigm and to fintech companies. The authors wish to thank
      Agostino Capponi, Mallesh Pai, and Anthony Zhang for helpful comments.}}
  \author{
    Austin Adams \\  Whetstone Research \\ email: \emailhref{austin@whetstone.cc} \and
    Ciamac C.\ Moallemi \\ Columbia University / Paradigm \\
    email: \emailhref{ciamac@gsb.columbia.edu}
    \and
    Sara Reynolds \\ Uniswap Labs \\ email: \emailhref{sara@uniswap.org} \and
    Dan Robinson \\ Paradigm \\ email: \emailhref{dan@paradigm.xyz}
  }
  \date{Initial version: February 13, 2024 \\ This version: February 12, 2025}
}

\begin{document}
\maketitle
\singlespacing

\begin{abstract}
  Automated market makers (AMMs) have emerged as the dominant market mechanism for trading on
  decentralized exchanges implemented on blockchains.
  This paper presents a single mechanism that targets two important unsolved problems for
  AMMs: reducing losses to informed orderflow, and maximizing revenue from
  uninformed orderflow. The ``auction-managed AMM'' works by running a censorship-resistant onchain
  auction for the right to temporarily act as ``pool manager'' for a constant-product AMM. The
  pool manager sets the swap fee rate on the pool, and also receives the accrued fees from
  swaps. The pool manager can exclusively capture some arbitrage by trading against the pool in
  response to small price movements, and also can set swap fees incorporating price sensitivity of
  retail orderflow and adapting to changing market conditions, with the benefits from both
  ultimately accruing to liquidity providers. Liquidity providers can enter and exit the pool
  freely in response to changing rent, though they must pay a small fee on withdrawal. We prove
  that under certain assumptions, this AMM should have higher liquidity in equilibrium than any
  standard, fixed-fee AMM.
\end{abstract}

%%% Local Variables:
%%% mode: latex
%%% TeX-master: "am-amm"
%%% End:

\iffalse
\begin{center}
  \textbf{\sffamily Preliminary version. Do not distribute.}
\end{center}
\fi

\onehalfspacing

\section{Introduction}

Liquidity providers (LPs) for automated market makers (AMMs) want to minimize their losses to
arbitrageurs while maximizing fee revenue from retail flow. Each of these is a major unsolved
problem in AMM design. Minimizing losses to arbitrageurs (which can be characterized as
``loss-vs-rebalancing,'' or LVR) requires either setting high fees or relying on some other
mechanism to capture the profits from information or latency arbitrage and mitigate losses to
agents (arbitrageurs), who possess superior information on the market value of the asset and snipe stale
prices posted by the AMM. Meanwhile, the optimal fee for a given asset pair depends on how retail
volume for that pair responds to fees --- a difficult problem that is not easy to model. Moreover,
the optimal fee may be dynamic and vary with other market variables such as volatility or overall
market volume. In standard fixed-fee AMMs (ff-AMMs), these problems compound and interfere with
each other: liquidity providers must choose a pool with fees high enough to reduce arbitrage
opportunities while still low enough to capture value from retail flow. Additionally, liquidity
providers must make this choice statically and for themselves, and if they disagree, liquidity
ends up fragmented across multiple pools.

% Benefits paragraph

In this paper, we propose the auction-managed AMM (am-AMM), a new AMM design that targets both LVR reduction and fee optimization with one mechanism. The mechanism incentivizes sophisticated market participants to capture some of the value leaked by the AMM, and also lets those market participants set fees at a level that optimizes revenue from retail traders. The pool maintains synchronous composability with other onchain contracts, does not require price oracles, and is resistant to censorship. It also ensures \textit{accessibility}, meaning that liquidity on the pool can always be traded against with some capped fee. Under certain assumptions, we prove that the am-AMM will attract more liquidity than \textit{any} fixed-fee constant product AMM pool (with any fee up to the cap), in equilibrium.

% How it works paragraph

The am-AMM is a constant-product AMM. The AMM utilizes an onchain censorship-resistant ``Harberger lease'' auction to find the highest bidder. The current highest bidder in the ongoing auction, known as the \textit{manager}, pays rent to liquidity providers. In exchange, the manager can dynamically set the swap fee rate (up to some maximum cap), and receives all swap fees collected by the pool. This allows the manager to capture small arbitrage opportunities each block by trading on the pool, since they can trade with effectively zero fee. It also incentivizes them to set the swap fee in order to maximize revenue from uninformed flow. Liquidity providers can enter and exit the pool freely (though they pay a small withdrawal fee).

The am-AMM has some drawbacks. It may provide even fewer protections against sandwich attacks than
a fixed-fee AMM, since the pool manager's ability to trade without fees allows them to profit by
pushing any publicly visible transaction to its limit price. It also could contribute to
centralization of block builder infrastructure. We think these limitations and drawbacks warrant
future study, as does the problem of making this feature work with concentrated liquidity
%\citep[see][]{Adams21}.
\cite{Adams21}.

\paraheader{Related literature.} While automated market makers are newer compared to many
financial market primitives, the literature on the subject is growing rapidly. The concept of AMMs
can be traced back to \citet{hanson2007logarithmic} and \citet{othman2013practical}. Early
literature on the current implementations of AMMs includes \citet{angeris2019analysis},
\citet{angeris2020improved}, \citet{lehar2021decentralized}, \citet{capponi2021adoption}, and
\citet{hasbrouck2022need}. Implementation details of automated market makers are described in
\citet{Adams20} and \citet{Adams21}. Furthermore, \citet{Adams23} describes a forthcoming platform
for customizable AMMs.

\todo{cite Canidio}
\todo{cite Nikete}
\todo{Conor McMenamin -- Diamond protocol}

Our paper builds directly on the loss-vs-rebalancing framework established in \citet{lvr2022} and
\citet{milionis2023automated} as a model for evaluating designs for AMMs. Contemporaneous with the
this paper, \citet{ma2024cost} analyze liquidity provision in AMMs with a similar model of
LP profitability and noise trader demand as the present paper.

The idea described in this paper is an instance of an \emph{ex ante} auction for the right to
capture the arbitrage profit from the block, an idea first proposed by Alex Herrmann from Gnosis
as the ``MEV capturing AMM'' (McAMM) in \citet{herrmann22mcamm}. This paper extends that concept
to allow the manager to also select the fee charged to retail traders and collect those fees, thus
using a similar mechanism to address an independent problem. Also, unlike the McAMM, the auction
proposed here guarantees the property of \textit{accessibility} --- people can trade against the
pool even if the current manager has not submitted a transaction in this block.

Our paper also proposes a way to optimize fees that takes into account the difficult-to-model
demand function for retail orderflow. Some AMMs, such as Uniswap v3 \citep{Adams21} address this
problem by letting liquidity providers choose between different static ``fee tiers.'' This puts
the responsibility for optimizing fees on individual liquidity providers, can lead to some
fragmentation of liquidity. Other automated market makers have implemented dynamic fees on
individual pools, including Trader Joe v2.1 \citep{mountainfarmer22joe}, Curve v2
\citep{egorov21curvev2}, and Mooniswap \citep{bukov20mooniswap}, as well as
\citet{Nezlobin2023}. \citet{cartea2023automated} propose a design involving dynamic fees and
dynamic price impact functions. While innovative, these dynamic fee implementations are not
necessarily optimized for maximizing liquidity. The am-AMM is a first attempt at a provably
incentive-compatible dynamic fee, where the fee is set by the market in a way that should always
attract more liquidity than any fixed-fee AMM.

%%% Local Variables:
%%% mode: latex
%%% TeX-master: "am-amm"
%%% End:

\section{Auction Design}

The auction is designed to set up a censorship-resistant version of the two-stage game modelled in
\Cref{sec:am-amm}, where (1) potential managers bid for the right to set and earn the trading fee in a future block; and (2) liquidity providers respond by adding or
removing liquidity. The auction uses a delay parameter $K$. The rules of the auction are designed
so that both the rent and the pool manager for block $N$ are locked in as of block $N - K$.

%\smallskip%
\paraheader{Harberger lease.} The right to be the manager of a pool is set in a
special onchain auction we call a ``Harberger lease'' (named after the ``Harberger taxes'' popularized by \citet{posner2018radical}). This is a continuously held English auction
where bids are expressed in terms of rent per block. The \textit{top bid} at any given time
determines the manager for the pool, who pays the rent to liquidity providers in the pool for as
long as that bid is active.

The rent is denominated and paid in pool tokens, and is paid out of the pool manager's deposit to
all pool LP token holders proportional to their stake (thus effectively increasing the value of those
pool tokens). This ensures that pool tokens remain fungible and that rent is compounded
automatically.

%\smallskip
\paraheader{Bidding rules.} Each bid specifies a per-block rent $R$. When the bid is placed, it must include a \textit{deposit} $D$, which must be a multiple of $R$ and must be at least $R \cdot K$. Newly placed bids do not become active immediately, but are delayed by $K$ blocks. When either a new high bid becomes active, or the current pool manager's deposit is depleted, the new high bidder \textit{usurps} the current pool manager. The contract enforces a minimum bid increment.

The top bid cannot be cancelled, but can reduce its deposit as long as it leave a minimum deposit of $D_{\mathrm{top}} \geq R_{\mathrm{top}} K$. The contract keeps track of the \textit{next} bid (which is either the next-highest bid after the current top bid, or a bid that is higher than the top bid but is not yet active). If the top bid does not have enough rent to pay for $K$ blocks --- that is, if $\frac{D_{\mathrm{top}}}{R_{\mathrm{top}}} < K$, then the next bid cannot be canceled, but must leave at least enough deposit such that $\frac{D_{\mathrm{top}}}{R_{\mathrm{top}}} + \frac{D_{\mathrm{next}}}{R_{\mathrm{next}}} \geq K$.

%\smallskip
\paraheader{Pool manager rights.} The current pool manager can adjust the swap fee for the next
block at any time, subject to a fixed \emph{fee cap} $f_{\max}$. The pool manager also receives all swap
fees collected by the pool. Since the pool manager receives swap fees, they are effectively able
to swap on the pool with zero fee. This means that they can capture arbitrages from small price
movements that no other arbitrageur would be able to profitably capture. However, if the price
moves by more than the current fee in a single block --- in other words, if it moves outside the
``no-trade region'' \citep{milionis2023automated} --- then some of the profit could leak to
other arbitrageurs, as discussed in \Cref{sec:am-amm}.

%\smallskip
\paraheader{Censorship resistance.} The delay parameter $K$ should be chosen such that there is a negligible probability that anyone can censor the base layer for $K$ blocks in a row. This prevents someone from stealing MEV from a high-volatility block by usurping the current manager and then censoring arbitrage transactions for $K$ blocks. It also ensures that liquidity providers can respond to changes in rent by adding or removing liquidity before the new rent takes effect.

%\smallskip
\paraheader{Withdrawal fees.} Liquidity providers are free to enter or exit at any time. This
allows them to respond to anticipated changes in rent during the $K$-block delay. However, when
liquidity providers exit, they must pay a small withdrawal fee to the current manager. This
prevents liquidity providers from withdrawing liquidity after volatility is realized
but before the current manager has the opportunity to execute an arbitrage transaction. As shown
in Appendix \ref{appendix:withdrawal}, even a very small withdrawal fee --- less than 0.13 basis
points, if the fee cap is 1\% --- can ensure that strategic liquidity provider withdrawals cannot
reduce the manager's profit from a given arbitrage opportunity below the amount assumed in
\Cref{theory}. This fee could alternatively be replaced with a withdrawal delay for a similar effect.

%%% Local Variables:
%%% mode: latex
%%% TeX-master: "am-amm"
%%% End:

\section{Theory} \label{theory}

Our starting point is a model inspired by \citet{lvr2022} and \citet{milionis2023automated}, which
consider an AMM trading a risky asset (denoted by $x$) versus the num\'eraire (denoted by $y$), and that the risky asset has a fundamental price at all times (for example, on an infinitely deep centralized exchange). As in \citet{milionis2023automated}, we assume that traders can only trade on the pool at discrete block generation times.

Our setting is consistent with and can be structurally microfounded in the full setting of \citet{milionis2023automated}, which assumes that blocks arrival times follow a Poisson process and that the asset's price follows geometric Brownian motion parameterized by volatility $\sigma > 0$, but our theorems do not depend on those assumptions. Instead we describe the model
primitives in %more abstract,
reduced form, with weaker assumptions, and provide examples that show that those models would satisfy the assumptions.

We will restrict to the case of a constant product market maker,\footnote{The key property of the
  constant product market maker we use is that the arbitrage profits and arbitrage excess scale
  linearly with pool value, where the constant of proportionality does not depend on the
  price. This property holds more generally for geometric mean market makers, and our results
  would trivially extend there. Beyond that, when considering more general invariant curves, there
  may be additional second order effects due to the price changing over time horizon of the LP
  investment. If price movements over the scale of the LP investment horizon are not large, these
  effects may not be significant, and we would expect the high level insights of our model to
  continue to apply to general invariant curves.}
with invariant $\sqrt{xy} = L$, where we denote
the reserve quantities by $(x,y)$ and the pool liquidity level by $L$. If the price of the risky
asset is given by $P$, the value of the the pool reserves is given by
% $V(L) = 2 \sqrt{P} L = V_0 L$, where $V_0 \defeq 2 \sqrt{P}$, as a function of the available
$V(L) \defeq 2 \sqrt{P} L$, as a function of the available
liquidity $L$. We assume the pool charges a proportional trading fee $f \in [0, f_{\max}]$.

We consider a setting where there are two types of traders: (1) noise traders, who trade for
idiosyncratic reasons and generate fee income for the pool; and (2) arbitrageurs, who seek to
exploit price differences between the pool and the fundamental price, and create adverse selection
costs for the pool. Here, the \pnl of LPs is a jump process, stochastically jumping at instances
of block generation. We will consider the expected aggregate instantaneous rate of \pnl of LPs per
unit time, where we are averaging over stochasticity in future price changes, or, equivalently,
assuming that exposure to market risk of the risky asset has been hedged.

%\smallskip
\paraheader{Noise traders.} We assume there exists a population of noise traders that trade for
idiosyncratic reasons (e.g., convenience of executing on chain) and not for informational
reasons. Hence, economically, in our model noise traders serve exclusively to generate fee income
for the pool. Given pool fee $f \geq 0$ and liquidity $L \geq 0$, denote by $H(f,L)$ the expected
total volume of noise trades arriving per unit time, denominated in the  num\'eraire, so that $f
H(f,L)$ is the total rate per unit time of fee revenue generated by the noise traders.
We will assume that:
\begin{assumption}[Noise trader demand]\label{as:noise-trader}
  For $L > 0$, define \[H_0(f,L) \defeq H(f,L) / V(L),\] to be the expected noise trader volume per
  unit time per unit of pool value, we assume that $H_0(\cdot,\cdot)$ is a continuous function
  satisfying:
  %\todo{fix}
  \begin{enumerate}
  \item\label{pt:nt-1} For all $L > 0$, $H_0(f,L)$ is a decreasing function of $f$.
    % Further, $f \mapsto f H_0(f,L)$ is a strictly concave function.
    % Further, $f \mapsto f H_0(f,L)$ is strictly concave, i.e., regular demand
    % from auction theory, \citep[see, e.g.][]{talluri2005theory}.
  \item\label{pt:nt-2} For all $f \in [0,f_{\max}]$, $H_0(f,L)$ is strictly decreasing function of
    $L$. Moreover, $H_0(f,L) \downarrow 0$ as $L \tends \infty$, and
    $H_0(f,L) \uparrow \infty$ as $L \tends 0$.
  \end{enumerate}
\end{assumption}
Part~\ref{pt:nt-1} asserts that noise trader demand is decreasing in the price (fee)
charged.
%Part~\ref{pt:nt-1} is a standard demand modeling assumption in revenue management \citep[see,
%e.g.][]{talluri2005theory}, it asserts that noise trader demand is decreasing in the price (fee)
%charged.
%, and that the marginal revenue from noise traders is also decreasing in
%price.
Part~\ref{pt:nt-2} implies that the noise trader is sub-linear in pool value or
liquidity, i.e., noise trader demand is satiated.

\begin{example}
  Consider $H(f,L) \defeq c_0 L^\alpha \exp(-c_1 f)$, where $c_0,c_1 > 0$ and $\alpha \in (0,1)$
  are parameters that could be estimated from transaction data, and may be time varying or may
  depend on other market parameters such as volatility or broader market volume. This liquidity
  dependence is consistent with the model proposed by \citet{Hasbrouck2023}.
\end{example}

%\smallskip%
\paraheader{Arbitrageur profits.} We assume there is a competitive and deep market of arbitrageurs
monitoring prices in the AMM and exploiting price differences between the AMM and the fundamental
price. Denote by $\ARBPROFIT(f,L)$ the expected profit to arbitrageurs per unit time, given fee
$f \in [0,f_{\max}]$ and liquidity $L \geq 0$.  We make the following assumption:
\begin{assumption}[Arbitrageur profits]\label{as:arb-profits}
  For $L > 0$, we assume that \[\ARBPROFIT(f,L) = \AP_0(f) V(L),\] where $\AP_0(\cdot)$ is a continuous,
  decreasing function.
\end{assumption}
\Cref{as:arb-profits} guarantees that arbitrage profits scale \emph{linearly} with the pool value
or liquidity, in contrast to noise trader volume. This is because we assume, by nature of engaging
in riskless activity, arbs have access to infinite capital for arbitrage activity, and arbitrage
demand will not be satiated so long as arbitrage opportunities are
available. \Cref{as:arb-profits} also implies that, holding liquidity fixed, arbitrageur profits are decreasing in the fee $f$, this
is because the fee is a friction that limits arbitrage.

This assumption does not depend on a particular model for either the asset price's behavior or for block generation times, but we show that it can be structurally microfounded in one such model, as illustrated here:\footnote{See
  \Cref{app:structural} for further discussion.}
\begin{example}\label{ex:arb-profits}
  Consider a setting where the risky asset's price follows geometric Brownian motion parameterized by volatility $\sigma > 0$ and blocks are generated at the arrivals of a Poisson process of rate
  $\Delta t^{-1}$, with $\Delta t$ being the average interblock time. \citet{milionis2023automated}
  establish that, when $\Delta t < 8\sigma^2$,
  \begin{equation}\label{eq:arb-profits}
    \ARBPROFIT(f,L) = \frac{\sigma^2}{8}
    \frac{1}{1 + \frac{f}{\sigma \sqrt{\Delta
          t/2}}}
    \frac{ e^{+f/2} +  e^{-f/2} }{2
      \parens[\Big]{ 1 - \sigma^2 \Delta t/ 8}}
    V(L).
  \end{equation}
  This satisfies \Cref{as:arb-profits}.
\end{example}

\subsection{Fixed-Fee AMM Model}

As a benchmark, we consider the standard AMM design with a fixed fee $f \in [0,f_{\max}]$, which
we refer to as a \emph{fixed-fee AMM} (ff-AMM). Following the discussion above, we define the
instantaneous rate of aggregate expected \pnl of LPs in this pool in excess of the risk free rate
according to
\begin{equation}\label{eq:pnlff}
  \begin{split}
    \PNLFF(f,L) & \defeq f H(f,L) - \ARBPROFIT(f,L)  - r V(L) \\
                & = \left(  f H_0(f,L) - \AP_0(f) - r  \right) V(L),
  \end{split}
\end{equation}
where the first term in the sum is the revenue from noise traders, the second term is the loss to
arbs, and the final term is a capital charge, where $r > 0$ is the risk-free rate.
Under free entry and exit of LPs, we define the
equilibrium in the fixed-fee AMM according to a zero profit condition:
\begin{lemma}[ff-AMM equilibrium]\label{lem:fixed-fee-eq}
  Given fee $f \in [0,f_{\max}]$, define a liquidity level $L^* > 0$ to be a competitive equilibrium if LPs earn
  zero profit in excess of the risk free rate, i.e., $\PNLFF(f,L^*) = 0$.
  Then, for any $f \in [0,f_{\max}]$, a unique equilibrium level of liquidity $L^*=\LFF(f)$ exists.
\end{lemma}

\subsection{Auction-Managed AMM Model}\label{sec:am-amm}

In this section, we consider the auction-managed AMM design.

%\smallskip%
\paraheader{Arbitrageur excess.}  In the am-AMM, the pool manager collects all of the fee
revenue, and therefore, effectively, can also trade against the pool without paying any
fees. Therefore, the pool will suffer adverse selection costs of
$\ARBPROFIT(0,L)$, independent of the fee level $f$. However, the manager does not
collect all of these fees as income. Instead, note that whenever the mispricing in the pool
exceeds the fee $f$, some of that mispricing can be profitably captured by agents other than the manager.
We denote by $\ARBEXCESS(f,L)$ the instantaneous rate of
expected arbitrage profit per unit time forgone by the manager. We assume that:
\begin{assumption}[Arbitrageur excess]\label{as:arb-excess}
  For $L > 0$, we assume that \[\ARBEXCESS(f,L) = \AE_0(f) V(L),\] with $\AE_0(f)$ a continuous,
  decreasing function that satisfies $\AE_0(f) \leq \AP_0(f)$, for all $f \in [0,f_{\max}]$, and
  where the inequality is strict if $f > 0$.
\end{assumption}
\Cref{as:arb-excess} is largely analogous to \Cref{as:arb-profits}, and can be similarly microfounded by a
stochastic model, as shown in \Cref{ex:arb-excess}, but does not depend on that model. The strict
inequality assumption simply asserts that \emph{some} of the arbitrage profit is captured by the
manager.

\begin{example}
  One extreme setting is the McAMM design of \citet{herrmann22mcamm}. There, no
  trades can occur in an AMM in a given block unless the pool manager has a transaction earlier in
  the block to ``unlock'' the pool. Hence the pool manager is guaranteed to be the first
  transaction in everyblock and can capture all arbitrage profits. In this case,
  $\ARBEXCESS(f,L)=0$.
\end{example}

\begin{example}\label{ex:arb-excess}
  In \Cref{app:structural}, we develop structural microfoundations of a model for arbitrageur excess in
  the setting of \citet{milionis2023automated}, where the risky asset's price follows geometric Brownian motion parameterized by volatility $\sigma > 0$ and blocks are generated at the arrivals of a Poisson process of rate $\Delta t^{-1}$, with $\Delta t$ being the average interblock time. In
  that setting, we assume that (1) the pool manager fully monetizes any arbitrage
  where the mispricing is less than the fee $f$, and (2) when the mispricing exceeds the fee
  $f$, other arbitrageurs correct the mispricing back to the fee level $f$ before the pool
  manager can trade, and hence the pool manager is only able to monetize the portion of the
  arbitrage up to $f$.
  Then, we establish that, when $\Delta t < 8\sigma^2$,
  \begin{equation}\label{eq:arb-excess}
    \ARBEXCESS(f,L) = \frac{\sigma^2}{8} \exp\left( - \frac{f}{\sigma \sqrt{\Delta t/2}}
    \right)
    \frac{ e^{+f/2} +  e^{-f/2} }{2
      \parens[\Big]{ 1 - \sigma^2 \Delta t/ 8}} V(L).
  \end{equation}
  This satisfies the conditions of \Cref{as:arb-excess}.
\end{example}

Comparing \Cref{ex:arb-profits} and \Cref{ex:arb-excess}, we see that $\ARBEXCESS(f,L) \ll
\ARBPROFIT(f,L)$ in the sense that
\[
  \frac{\ARBEXCESS(f,L)}{\ARBPROFIT(f,L)}
  =
  \frac{\AE_0(f,L)}{\AP_0(f,L)}
  =
  \left( 1 + \frac{f}{\sigma \sqrt{\Delta
        t/2}} \right)
  \exp\left( - \frac{f}{\sigma \sqrt{\Delta t/2}}\right),
\]
which is exponentially vanishing in $f/\sigma\sqrt{\Delta t}$.

%\smallskip
\paraheader{Equilibrium.}
We imagine a game that proceeds in two steps: (1) agents bid rent $R$, the agent with the
highest rent wins the auction and is declared the pool manager, with the right to determine the
trading fee $f$ and earn all fees; and
(2) LPs determine aggregate liquidity $L$ given $R$. Based on the discussion above,
the \pnl of the manager per unit time is given by
\[
  \begin{split}
    \PNLMGR(R,L) & \defeq \max_{f\in[0,f_{\max}]}\ \left\{ f H(f,L) + \ARBPROFIT(0,L)
                   \right. \\
    & \quad\quad\quad\quad\quad\quad \left.
                   -
                 \ARBEXCESS(f,L) - R \right\} \\
    & = \max_{f\in[0,f_{\max}]}\ \left\{ f H_0(f,L) + \AP_0(0) -
      \AE_0(f) \right\} V(L) - R.
  \end{split}
\]
The maximization is because the manager is free to set the fee, and we assume they will do so to
maximize \pnl. The first term captures the fact that the manager retains all fee revenue. The second and third
terms capture the fact that the manager earns arbitrage profits as if it pays no fees, except
for the arbitrage excess. The final term captures the fact that the manager pays rent. On the other
hand, the LPs in aggregate earn \pnl per unit time given by
\[
  \PNLLP(R,L) \defeq R - \ARBPROFIT(0,L)  - r V(L) = R - \left( \AP_0(0) + r \right) V(L).
\]
Here, the first term is the rent payment made by the manager, the second term is the adverse
selection cost (which is as if no fees are charged), and the third term is the cost of capital.
Given these definitions, we have that:
\begin{theorem}[am-AMM equilibrium]\label{th:liq-dom}
  $(R^*,L^*)$ is a competitive equilibrium of the am-AMM if the zero profit
  conditions\footnote{Here, we assume that entry and exit are frictionless for LPs, ignoring the
    withdrawal fees. This is justified if the liquidity provision decision horizon is sufficiently
    long so that, amortized over its length, the withdrawal fees are de minimus.}
  \[
    \PNLMGR(R^*,L^*)=0,\qquad \PNLLP(R^*,L^*)=0,
  \]
  are satisfied. An equilibrium $(R^*,L^*)$ must
  exist, with equilibrium fees
  \begin{equation}\label{eq:am-amm-fees}
    f^* \in \argmax_{f\in[0,f_{\max}]}\ f  H_0(f,L) - \AE_0(f),
  \end{equation}
  and satisfies $L^* > \LFF(f)$, for all $f \geq 0$. Therefore, in equilibrium, the am-AMM will
  have higher liquidity than any ff-AMM.
\end{theorem}

\Cref{th:liq-dom} establishes that the equilibrium liquidity for the am-AMM is larger than that of
any ff-AMM. Beyond that, it gives insight into the equilibrium fee $f^*$ set in the am-AMM: from
\eqref{eq:am-amm-fees}, the pool operator will seek to set the fee to maximize noise trader
revenue adjusted for arbitrageur excess. As a comparison, consider the fee level \fopt that exclusively
maximizes noise trader revenue, i.e.,
\begin{equation}\label{eq:opt-fees}
  \fopt \in \argmax_{f\in[0,f_{\max}]}\ f  H_0(f,L^*).
\end{equation}
If we assume that the revenue function $f \mapsto f H_0(f,L^*)$ is concave, and the arbitrageur excess
function $\AE_0(f)$ is convex, then a comparison of first order conditions for
\eqref{eq:am-amm-fees}--\eqref{eq:opt-fees} reveals that $\fopt \leq f^*$, i.e., the am-AMM will
set fees higher than is purely revenue optimal. However, by virtue of the optimality of $f^*$ in
\eqref{eq:am-amm-fees}, we have that
\begin{equation}\label{eq:opt-fees2}
  \fopt H_0(\fopt, L^*) - f^* H_0(f^*,L^*) \leq \AE_0(\fopt) - \AE_0(f^*) \leq \AE_0(\fopt).
\end{equation}
In the arbitrageur excess model of \Cref{ex:arb-excess}, $\AE_0(\fopt)$ may be very small (since it is
exponentially vanishing in blocktime, for example), and in such cases \eqref{eq:opt-fees2} would
imply that $f^*$ is also nearly optimal from a noise trader revenue perspective.

%%% Local Variables:
%%% mode: latex
%%% TeX-master: "am-amm"
%%% End:

\section{A Structural Model for Arbitrageur  Excess}\label{app:structural}

In this section, we will derive the structural model for arbitrageur excess of \Cref{ex:arb-excess},
following the arbitrageur profits model of \citet{milionis2023automated} and of \Cref{ex:arb-profits}.

\paraheader{Fee structure.}  In order to simplify formulas, \citet{milionis2023automated} uses a
fee structure wherein a proportional fee of $e^{+\gamma}-1 = \gamma + o(\gamma)$ is charged for
purchases of the risky asset from the pool, while a proportional fee of
$1-e^{-\gamma} = \gamma + o(\gamma)$ is charged for sales of the risky asset to the pool --- these
are symmetric fees in log-price space. This is different than the setting in this paper, where we
assume a fee proportional $f$ which is the same for buys or sells --- that is, symmetric in
(linear) price space. In order to facilitate comparison with \citet{milionis2023automated}, we
will make the approximation $f = \gamma$. As illustrated in Example~1 of
\citet{milionis2023automated}, for practical parameter values, this does not make a significant
difference.

\paraheader{Asset price dynamics.}
Denote by $P_t$ the fundamental price of the asset, which follows a geometric Brownian motion with
drift $\mu > 0$ and volatity $\sigma > 0$, and denote by $\tilde P_t$ the implied spot price of
the asset in the pool. Define $z_t \defeq \log P_t/\tilde P_t$ to be the log-mispricing at time
$t$. When $t$ is not a block generation time,  It\^{o}'s lemma implies that $z_t$ is governed by
the stochastic differential equation
\begin{equation}\label{eq:z-dyn2}
  dz_t = \left( \mu - \tfrac{1}{2} \sigma^2\right) \, dt
  + \sigma\, dB_t,
\end{equation}
where $B_t$ is a Brownian motion. We will make the symmetry assumption that $\mu =
\tfrac{1}{2} \sigma^2$, so that $z_t$ is a driftless, (scaled) Brownian motion.

\paraheader{Block time dynamics.}
We assume that blocks are generated according to a Poisson
process with mean interarrival time $\Delta t$. Denote the block generation times by
$0< \tau_1 < \tau_2 < \ldots$. At instances $t=\tau_i$ when blocks are generated, we imagine that
\begin{enumerate}
\item If the $|z_{t-}| \geq f$ (i.e., the mispricing immediately before block generation
  exceeds the fee), we imagine an arbitrageur is able to trade until the mispricing is equal to the fee,
  and thus earns profits that are not captured by the pool operator. These profits then become
  part of arbitrageur excess. Following the derivation of \citet{milionis2023automated},
  the instaneous
  profit from arbitrageur excess when price is $P=P_t$ and the mispricing is $z=z_{t-}$
  due to buying (respectively, selling) is given by
  \[
    A_+(P,z) \defeq
    \left[
      P
      \left\{ x^*\left(P e^{-z} \vphantom{\tilde P} \right)
        - x^*\left(P e^{-f}\right)  \right\}
      + e^{+f}
      \left\{y^*\left(P e^{-z} \vphantom{\tilde P} \right)
        -  y^*\left(P e^{-f} \right) \right\}
    \right] \I{ z > +f } \geq 0,
  \]
  \[
    A_-(P,z) \defeq
    \left[
      P
      \left\{ x^*\left(P e^{-z} \vphantom{\tilde P} \right)
        - x^*\left(P e^{+f} \vphantom{\tilde P} \right)  \right\}
      + e^{-f} \left\{ y^*\left(P e^{-z} \vphantom{\tilde P} \right)
        -  y^*\left(P e^{+f} \vphantom{\tilde P} \right) \right\}
    \right] \I{ z < -f } \geq 0.
  \]
  Here, the holdings of the pool when implied price is $P$ are given by $x^*(P) \defeq
  L/\sqrt{P}$, $y^*(P) = L \sqrt{P}$.
\item After the arbitrageur trades, the pool operator corrects any remaining mispricing, so that $z_t=0$.
\end{enumerate}

\paraheader{Intensity of arbitrageur excess.}
The intensity or instaneous rate of arbitrageur excess per dollar of pool value per unit time is given by
\[
  \AE_0(f) = \frac{\ARBEXCESS(f,L)}{V(L)} =
  \frac{1}{\Delta t}\E\left[
\frac{  A_+(P,z) +   A_-(P,z) }{V(L)}
  \right],
\]
where the expectation is over $z=z_\tau$ at the next block generation time $\tau$, i.e., $z\sim
N(0,\sigma^2 \tau)$ with $\tau \sim \text{Exp}(\Delta t^{-1})$.

Observe that
\[
  \begin{split}
    \frac{A_+(P,z)}{V(L)}
    & =
      \frac{1}{2 L \sqrt{P}}
      \left[
      P
      \left\{ x^*\left(P e^{-z} \vphantom{\tilde P} \right)
      - x^*\left(P e^{-f}\right)  \right\}
      + e^{+f}
      \left\{y^*\left(P e^{-z} \vphantom{\tilde P} \right)
      -  y^*\left(P e^{-f} \right) \right\}
      \right] \I{ z > +f }
    \\
    & = \tfrac{1}{2}
      \left[
      \left\{
      e^{+z/2} -  e^{+f/2}
      \right\}
      + e^{+f}
      \left\{e^{-z/2} -  e^{-f/2}  \right\}
      \right] \I{ z > +f }
    \\
    & = \tfrac{1}{2}
      e^{+f/2}
      \left[
      e^{+(z-f)/2}
      -  2
      +
      e^{-(z-f)/2}
      \right] \I{ z > +f },
    \\
    \frac{A_-(P,z)}{V(L)}
    & =
      \frac{1}{2 L \sqrt{P}}
      \left[
      P
      \left\{ x^*\left(P e^{-z} \vphantom{\tilde P} \right)
      - x^*\left(P e^{+f}\right)  \right\}
      + e^{-f}
      \left\{y^*\left(P e^{-z} \vphantom{\tilde P} \right)
      -  y^*\left(P e^{+f} \right) \right\}
      \right] \I{ z < -f }
    \\
    & = \tfrac{1}{2}
      \left[
      \left\{
      e^{+z/2} -  e^{-f/2}
      \right\}
      + e^{-f}
      \left\{e^{-z/2} -  e^{+f/2}  \right\}
      \right] \I{ z < -f }
    \\
    & = \tfrac{1}{2}
      e^{-f/2}
      \left[
      e^{+(z+f)/2}
      -  2
      +
      e^{-(z+f)/2}
      \right] \I{ z < -f },
  \end{split}
\]

% reference: https://www.desmos.com/calculator/dumjqgbrkb

Taking expectations of the first term, conditioned on the block generation time $\tau$, we have
$z\sim N(0,\sigma^2 \tau)$, so that
\[
  \begin{split}
    \E\bracks*{\left. \frac{ A_+(P,z)  }{V(L)} \right|\ \tau}
    & =
      -e^{\frac{f}{2}}+\frac{1}{2}\left(e^{f}+1\right)e^{\frac{\sigma^{2}\tau}{8}}-\frac{1}{2}e^{f}e^{\frac{\sigma^{2}\tau}{8}}\operatorname{erf}\left(\frac{\sigma^{2}\tau+2f}{2\sqrt{2}\sigma\sqrt{\tau}}\right)
    \\
    & \quad +\frac{1}{2}e^{\frac{\sigma^{2}\tau}{8}}\operatorname{erf}\left(\frac{\sigma^{2}\tau-2f}{2\sqrt{2}\sigma\sqrt{\tau}}\right)+e^{\frac{f}{2}}\operatorname{erf}\left(\frac{f}{\sigma\sqrt{\tau}\sqrt{2}}\right).
  \end{split}
\]
Taking expectations over $\tau \sim \text{Exp}(\Delta t^{-1})$, assuming that $\Delta t < 8\sigma^2$,
\[
  \begin{split}
    \frac{1}{\Delta t} \E\bracks*{ \frac{ A_+(P,z)  }{V(L)} }
    & =
      \frac{\sigma^2}{8} \exp\left( - \frac{f}{\sigma \sqrt{\Delta t/2}}
    \right)
    \frac{ e^{+f/2} }{2
      \parens[\Big]{ 1 - \sigma^2 \Delta t/ 8}}.
  \end{split}
\]
Similarly, for the other term,
\[
  \begin{split}
    \frac{1}{\Delta t} \E\bracks*{ \frac{ A_-(P,z)  }{V(L)} }
    & =
      \frac{\sigma^2}{8}
      \exp\left( - \frac{f}{\sigma \sqrt{\Delta t/2}}
      \right)
    \frac{ e^{-f/2} }{2
      \parens[\Big]{ 1 - \sigma^2 \Delta t/ 8}}.
  \end{split}
\]
Combining these results yields the arbitrageur excess expression in \eqref{eq:arb-excess}.

\paraheader{Discussion.}
The expressions \eqref{eq:arb-profits} for arbitrageur profits and \eqref{eq:arb-excess} for arb
excess have an interesting common structure. In both cases, the expressions can be decomposed into
the product of (1) the probability that an arbitrageur can trade at the next block time; and (2) the
expected profit conditioned on trade. In the arbitrageur profits expression, the probability of trade is
given by
\[
  \frac{1}{1 + \frac{f}{\sigma \sqrt{\Delta
        t/2}}},
\]
while in the arbitrageur excess expression, it is given by
\[
  \exp\left( - \frac{f}{\sigma \sqrt{\Delta t/2}}
  \right).
\]
In both cases, however, the expected profit conditioned on trade is the same. Thus, the fact that
arbitrageur excess is less than arbitrageur profits is driven by the fact that the probability of trade is
(exponentially) less. This, in turn, is because the pool operator drives the mispricing to zero at
the end of every block, making large mispricings at the beginning of the next block unlikely.

%%% Local Variables:
%%% mode: latex
%%% TeX-master: "am-amm"
%%% End:

\section{Discussion}

\paraheader{Advantages.}
One desirable characteristic of the auction-managed AMM is that it shifts the strategic burden of
determining the optimal fee from passive LPs to the pool manager, who, in turn will set the fee
according to \eqref{eq:am-amm-fees}, which seeks to set fees to optimize revenue from noise
traders, adjusted by arb excess paid to arbitraguers. This makes sense, since the pool manager is
assumed to be a more sophisticated entity that can perform off-chain modeling and analysis to
estimate noise trader sensitivity and losses to arbs.

Our model captures the most visible and salient features that drive noise trader demand (the fee
and the liquidity). However, as argued by \citet{milionis2023automated} and
\citet{trianglefees2023}, noise traders may also be concerned with the accuracy of prices in the
pool. Relative to an outside reference price (e.g., the price on infinitely deep centralized
exchange), noise traders pay an effective spread which is the sum of the trading fee of the pool
(deterministic, positive) and the relative mispricing of the pool (stochastic, could be positive
or negative). In general noise traders will prefer pools with less relative mispricing.  An
additional benefit of the am-AMM over the ff-AMM is that the quoted prices are more accurate. This
is because the pool operator of an am-AMM faces no fees in performing arbitrage trades against the
pool, and is able to correct smaller price discrepancies.

There is also an important transfer of risk
between the LPs and the pool manager: in the am-AMM, the LPs earn rent payments instead of noise
trader fees. Since these rent payments are determined \emph{ex ante}, in equilibrium, they
incorporate the expected value of future noise trader fee revenue, in contrast to the actual noise
trader fee revenue, which arrives in a lumpy and stochastic fashion and goes to the pool
manager. Although it is beyond the scope of the analysis we have presented (which assumes risk
neutrality), this risk transfer is likely welfare improving since the pool manager is likely a larger
and better capitalized entity than a typical passive LP, and is thus less risk averse.

\paraheader{Drawbacks.}
The auction-managed AMM is not without drawbacks. First, the pool manager's ability to effectively
trade on the pool with zero spread exacerbates the ``sandwich attack'' problem with onchain AMMs,
as discussed generally by \citet{daian2020flash,zhou2021high,adams2023costs}. A
party that can trade with zero fees can profit by pushing any publicly visible swap transaction to
its limit price. This is a meaningful drawback — which could hurt swappers or ultimately
discourage retail flow, thus potentially invalidating the assumption in \Cref{theory} that retail flow is only a function of liquidity and fee --- but known mitigations for general sandwich attacks could be applied, such
as verifiable sequencing rules \citep{ferreira2022credible}, private relays
\citep{adams2023costs}, or offchain filler auctions \citep{adams2023uniswapx}. %We consider such
%specific mitigations out of scope for this paper.

Second, the \emph{ex ante} auction for the arbitrage opportunity could have effects on the market for block
building. The pool manager's exclusive right to capture some arbitrage profit from a pool
could give them an advantage in the block builder auction, similar to the advantages enjoyed by
builders with private orderflow discussed by \citet{gupta2023centralizing}. Additionally, while the
am-AMM allows any party to bid on the pool manager position, parties with greater sophistication
or access to private orderflow may be more likely to win the auction. On the other hand, by capturing
some arbitrage profit and reducing the power of block proposers, the am-AMM could mitigate some of the
harmful pressures that MEV puts on the blockchain consensus mechanism. We think these possible consequences 
warrant further study.

Lastly, one disadvantage of the am-AMM over the McAMM discussed by \citet{herrmann22mcamm} is that
the current manager does not capture the ``arbitrage excess'' that results from single-block price
movements beyond the no-trade region, as discussed in \Cref{theory}. This creates an
incentive for the manager to set a higher fee than the one that optimizes total fee revenue. This
could be fixed by requiring the manager to unlock the pool at every block, at the cost of
sacrificing the accessibility property.

%\smallskip%
\paraheader{Future work.}
In this paper, we focus on constant product market makers. Extending this design to other AMMs —
and particularly to concentrated liquidity AMMs, in which different liquidity providers may have
different returns during the same period and may go in or out of range as a result of price
changes — is left for future work.

Implementing this mechanism is also left for future work. Since we first made this paper available, there has been at least one open-source third-party implementation \cite{biddog}.

%This paper does not attempt to quantify the ``excess arbitrage'' that is leaked by the mechanism %when the price moves by more than the fee in a single block.

%%% Local Variables:
%%% mode: latex
%%% TeX-master: "am-amm"
%%% End:

{\small
  \singlespacing
  \bibliographystyle{ACM-Reference-Format}
  \bibliography{references}


\begin{thebibliography}{29}

%%% the Web address.

\ifx \showCODEN    \undefined \def \showCODEN     #1{\unskip}     \fi
\ifx \showDOI      \undefined \def \showDOI       #1{#1}\fi
\ifx \showISBNx    \undefined \def \showISBNx     #1{\unskip}     \fi
\ifx \showISBNxiii \undefined \def \showISBNxiii  #1{\unskip}     \fi
\ifx \showISSN     \undefined \def \showISSN      #1{\unskip}     \fi
\ifx \showLCCN     \undefined \def \showLCCN      #1{\unskip}     \fi
\ifx \shownote     \undefined \def \shownote      #1{#1}          \fi
\ifx \showarticletitle \undefined \def \showarticletitle #1{#1}   \fi
\ifx \showURL      \undefined \def \showURL       {\relax}        \fi
% invisible to TeX
\providecommand\bibfield[2]{#2}
\providecommand\bibinfo[2]{#2}
\providecommand\natexlab[1]{#1}
\providecommand\showeprint[2][]{arXiv:#2}

\bibitem[Adams et~al\mbox{.}(2023a)]%
        {adams2023costs}
\bibfield{author}{\bibinfo{person}{Austin Adams}, \bibinfo{person}{Benjamin~Y
  Chan}, \bibinfo{person}{Sarit Markovich}, {and} \bibinfo{person}{Xin Wan}.}
  \bibinfo{year}{2023}\natexlab{a}.
\newblock \showarticletitle{The Costs of Swapping on the Uniswap Protocol}.
\newblock \bibinfo{journal}{\emph{arXiv preprint arXiv:2309.13648}}
  (\bibinfo{year}{2023}).
\newblock


\bibitem[Adams et~al\mbox{.}(2023b)]%
        {Adams23}
\bibfield{author}{\bibinfo{person}{Hayden Adams}, \bibinfo{person}{Moody
  Salem}, \bibinfo{person}{Noah Zinsmeister}, \bibinfo{person}{Sara Reynolds},
  \bibinfo{person}{Austin Adams}, \bibinfo{person}{Will Pote},
  \bibinfo{person}{Mark Toda}, \bibinfo{person}{Alice Henshaw},
  \bibinfo{person}{Emily Williams}, {and} \bibinfo{person}{Dan Robinson}.}
  \bibinfo{year}{2023}\natexlab{b}.
\newblock \bibinfo{booktitle}{\emph{Uniswap v4 Core [Draft]}}.
\newblock
\urldef\tempurl%
\url{https://github.com/Uniswap/v4-core/blob/main/docs/whitepaper-v4.pdf}
\showURL{%
\tempurl}


\bibitem[Adams et~al\mbox{.}(2020)]%
        {Adams20}
\bibfield{author}{\bibinfo{person}{Hayden Adams}, \bibinfo{person}{Noah
  Zinsmeister}, {and} \bibinfo{person}{Dan Robinson}.}
  \bibinfo{year}{2020}\natexlab{}.
\newblock \bibinfo{booktitle}{\emph{Uniswap v2 Core}}.
\newblock
\urldef\tempurl%
\url{https://uniswap.org/whitepaper.pdf}
\showURL{%
Retrieved Jun 12, 2023 from \tempurl}


\bibitem[Adams et~al\mbox{.}(2021)]%
        {Adams21}
\bibfield{author}{\bibinfo{person}{Hayden Adams}, \bibinfo{person}{Noah
  Zinsmeister}, \bibinfo{person}{Moody Salem}, \bibinfo{person}{River Keefer},
  {and} \bibinfo{person}{Dan Robinson}.} \bibinfo{year}{2021}\natexlab{}.
\newblock \bibinfo{booktitle}{\emph{Uniswap v3 Core}}.
\newblock
\urldef\tempurl%
\url{https://uniswap.org/whitepaper-v3.pdf}
\showURL{%
Retrieved Jun 12, 2023 from \tempurl}


\bibitem[Adams et~al\mbox{.}(2023c)]%
        {adams2023uniswapx}
\bibfield{author}{\bibinfo{person}{Hayden Adams}, \bibinfo{person}{Noah
  Zinsmeister}, \bibinfo{person}{Mark Toda}, \bibinfo{person}{Emily Williams},
  \bibinfo{person}{Xin Wan}, \bibinfo{person}{Matteo Leibowitz},
  \bibinfo{person}{Will Pote}, \bibinfo{person}{Allen Lin},
  \bibinfo{person}{Eric Zhong}, \bibinfo{person}{Zhiyuan Yang},
  \bibinfo{person}{Riley Campbell}, \bibinfo{person}{Alex Karys}, {and}
  \bibinfo{person}{Dan Robinson}.} \bibinfo{year}{2023}\natexlab{c}.
\newblock \bibinfo{booktitle}{\emph{UniswapX}}.
\newblock
\urldef\tempurl%
\url{https://uniswap.org/whitepaper-uniswapx.pdf}
\showURL{%
\tempurl}


\bibitem[Angeris and Chitra(2020)]%
        {angeris2020improved}
\bibfield{author}{\bibinfo{person}{Guillermo Angeris} {and}
  \bibinfo{person}{Tarun Chitra}.} \bibinfo{year}{2020}\natexlab{}.
\newblock \showarticletitle{Improved price oracles: Constant function market
  makers}. In \bibinfo{booktitle}{\emph{Proceedings of the 2nd ACM Conference
  on Advances in Financial Technologies}}. \bibinfo{pages}{80--91}.
\newblock


\bibitem[Angeris et~al\mbox{.}(2019)]%
        {angeris2019analysis}
\bibfield{author}{\bibinfo{person}{Guillermo Angeris},
  \bibinfo{person}{Hsien-Tang Kao}, \bibinfo{person}{Rei Chiang},
  \bibinfo{person}{Charlie Noyes}, {and} \bibinfo{person}{Tarun Chitra}.}
  \bibinfo{year}{2019}\natexlab{}.
\newblock \showarticletitle{An analysis of Uniswap markets}.
\newblock \bibinfo{journal}{\emph{arXiv preprint arXiv:1911.03380}}
  (\bibinfo{year}{2019}).
\newblock


\bibitem[BidDog(2024)]%
        {biddog}
\bibfield{author}{\bibinfo{person}{BidDog}.} \bibinfo{year}{2024}\natexlab{}.
\newblock \bibinfo{title}{BidDog: Open source implementation of am-AMM
  auctions}.
\newblock
\newblock
\urldef\tempurl%
\url{https://github.com/Bunniapp/biddog}
\showURL{%
\tempurl}
\newblock
\shownote{Accessed: 2024-05-22}.


\bibitem[Bukov and Melnik(2020)]%
        {bukov20mooniswap}
\bibfield{author}{\bibinfo{person}{Anton Bukov} {and} \bibinfo{person}{Mikhail
  Melnik}.} \bibinfo{year}{2020}\natexlab{}.
\newblock \bibinfo{booktitle}{\emph{Mooniswap by 1inch.exchange}}.
\newblock
\urldef\tempurl%
\url{https://mooniswap.exchange/docs/MooniswapWhitePaper-v1.0.pdf}
\showURL{%
Retrieved Sept 18, 2023 from \tempurl}


\bibitem[Capponi and Jia(2021)]%
        {capponi2021adoption}
\bibfield{author}{\bibinfo{person}{Agostino Capponi} {and}
  \bibinfo{person}{Ruizhe Jia}.} \bibinfo{year}{2021}\natexlab{}.
\newblock \showarticletitle{The adoption of blockchain-based decentralized
  exchanges}.
\newblock \bibinfo{journal}{\emph{arXiv preprint arXiv:2103.08842}}
  (\bibinfo{year}{2021}).
\newblock


\bibitem[Cartea et~al\mbox{.}(2023)]%
        {cartea2023automated}
\bibfield{author}{\bibinfo{person}{{\'A}lvaro Cartea},
  \bibinfo{person}{Fay{\c{c}}al Drissi}, \bibinfo{person}{Leandro
  S{\'a}nchez-Betancourt}, \bibinfo{person}{David Siska}, {and}
  \bibinfo{person}{Lukasz Szpruch}.} \bibinfo{year}{2023}\natexlab{}.
\newblock \showarticletitle{Automated market makers designs beyond constant
  functions}.
\newblock \bibinfo{journal}{\emph{Available at SSRN 4459177}}
  (\bibinfo{year}{2023}).
\newblock


\bibitem[Daian et~al\mbox{.}(2020)]%
        {daian2020flash}
\bibfield{author}{\bibinfo{person}{Philip Daian}, \bibinfo{person}{Steven
  Goldfeder}, \bibinfo{person}{Tyler Kell}, \bibinfo{person}{Yunqi Li},
  \bibinfo{person}{Xueyuan Zhao}, \bibinfo{person}{Iddo Bentov},
  \bibinfo{person}{Lorenz Breidenbach}, {and} \bibinfo{person}{Ari Juels}.}
  \bibinfo{year}{2020}\natexlab{}.
\newblock \showarticletitle{Flash boys 2.0: Frontrunning in decentralized
  exchanges, miner extractable value, and consensus instability}. In
  \bibinfo{booktitle}{\emph{2020 IEEE Symposium on Security and Privacy (SP)}}.
  IEEE, \bibinfo{pages}{910--927}.
\newblock


\bibitem[Egorov and GmbH)(2021)]%
        {egorov21curvev2}
\bibfield{author}{\bibinfo{person}{Michael Egorov} {and} \bibinfo{person}{Curve
  Finance (Swiss~Stake GmbH)}.} \bibinfo{year}{2021}\natexlab{}.
\newblock \bibinfo{booktitle}{\emph{Automatic market-making with dynamic peg}}.
\newblock
\urldef\tempurl%
\url{https://classic.curve.fi/files/crypto-pools-paper.pdf}
\showURL{%
Retrieved Sept 18, 2023 from \tempurl}


\bibitem[Ferreira and Parkes(2022)]%
        {ferreira2022credible}
\bibfield{author}{\bibinfo{person}{Matheus~VX Ferreira} {and}
  \bibinfo{person}{David~C Parkes}.} \bibinfo{year}{2022}\natexlab{}.
\newblock \showarticletitle{Credible decentralized exchange design via
  verifiable sequencing rules}.
\newblock \bibinfo{journal}{\emph{arXiv preprint arXiv:2209.15569}}
  (\bibinfo{year}{2022}).
\newblock


\bibitem[Gupta et~al\mbox{.}(2023)]%
        {gupta2023centralizing}
\bibfield{author}{\bibinfo{person}{Tivas Gupta}, \bibinfo{person}{Mallesh~M
  Pai}, {and} \bibinfo{person}{Max Resnick}.} \bibinfo{year}{2023}\natexlab{}.
\newblock \showarticletitle{The centralizing effects of private order flow on
  proposer-builder separation}.
\newblock \bibinfo{journal}{\emph{arXiv preprint arXiv:2305.19150}}
  (\bibinfo{year}{2023}).
\newblock


\bibitem[Hanson(2007)]%
        {hanson2007logarithmic}
\bibfield{author}{\bibinfo{person}{Robin Hanson}.}
  \bibinfo{year}{2007}\natexlab{}.
\newblock \showarticletitle{Logarithmic markets coring rules for modular
  combinatorial information aggregation}.
\newblock \bibinfo{journal}{\emph{The Journal of Prediction Markets}}
  \bibinfo{volume}{1}, \bibinfo{number}{1} (\bibinfo{year}{2007}),
  \bibinfo{pages}{3--15}.
\newblock


\bibitem[Hasbrouck et~al\mbox{.}(2022)]%
        {hasbrouck2022need}
\bibfield{author}{\bibinfo{person}{Joel Hasbrouck}, \bibinfo{person}{Thomas~J
  Rivera}, {and} \bibinfo{person}{Fahad Saleh}.}
  \bibinfo{year}{2022}\natexlab{}.
\newblock \showarticletitle{The need for fees at a dex: How increases in fees
  can increase dex trading volume}.
\newblock \bibinfo{journal}{\emph{Available at SSRN}} (\bibinfo{year}{2022}).
\newblock


\bibitem[Hasbrouck et~al\mbox{.}(2023)]%
        {Hasbrouck2023}
\bibfield{author}{\bibinfo{person}{Joel Hasbrouck}, \bibinfo{person}{Thomas~J.
  Rivera}, {and} \bibinfo{person}{Fahad Saleh}.}
  \bibinfo{year}{2023}\natexlab{}.
\newblock \bibinfo{title}{An Economic Model of a Decentralized Exchange with
  Concentrated Liquidity}.  (\bibinfo{year}{2023}).
\newblock
\newblock
\shownote{Working paper}.


\bibitem[Herrmann(2022)]%
        {herrmann22mcamm}
\bibfield{author}{\bibinfo{person}{Alex Herrmann}.}
  \bibinfo{year}{2022}\natexlab{}.
\newblock \bibinfo{booktitle}{\emph{MEV capturing AMM (McAMM)}}.
\newblock
\urldef\tempurl%
\url{https://ethresear.ch/t/mev-capturing-amm-mcamm/13336}
\showURL{%
Retrieved Sept 18, 2023 from \tempurl}


\bibitem[Lehar and Parlour(2021)]%
        {lehar2021decentralized}
\bibfield{author}{\bibinfo{person}{Alfred Lehar} {and}
  \bibinfo{person}{Christine~A Parlour}.} \bibinfo{year}{2021}\natexlab{}.
\newblock \showarticletitle{Decentralized exchanges}.
\newblock \bibinfo{journal}{\emph{Available at SSRN 3905316}}
  (\bibinfo{year}{2021}).
\newblock


\bibitem[Ma and Crapis(2024)]%
        {ma2024cost}
\bibfield{author}{\bibinfo{person}{Julian Ma} {and} \bibinfo{person}{Davide
  Crapis}.} \bibinfo{year}{2024}\natexlab{}.
\newblock \showarticletitle{The Cost of Permissionless Liquidity Provision in
  Automated Market Makers}.
\newblock \bibinfo{journal}{\emph{arXiv preprint arXiv:2402.18256}}
  (\bibinfo{year}{2024}).
\newblock


\bibitem[Milionis et~al\mbox{.}(2023)]%
        {milionis2023automated}
\bibfield{author}{\bibinfo{person}{Jason Milionis}, \bibinfo{person}{Ciamac~C
  Moallemi}, {and} \bibinfo{person}{Tim Roughgarden}.}
  \bibinfo{year}{2023}\natexlab{}.
\newblock \showarticletitle{Automated Market Making and Arbitrage Profits in
  the Presence of Fees}.
\newblock \bibinfo{journal}{\emph{arXiv preprint arXiv:2305.14604}}
  (\bibinfo{year}{2023}).
\newblock


\bibitem[Milionis et~al\mbox{.}(2022)]%
        {lvr2022}
\bibfield{author}{\bibinfo{person}{Jason Milionis}, \bibinfo{person}{Ciamac~C.
  Moallemi}, \bibinfo{person}{Tim Roughgarden}, {and}
  \bibinfo{person}{Anthony~Lee Zhang}.} \bibinfo{year}{2022}\natexlab{}.
\newblock \bibinfo{title}{Automated Market Making and Loss-Versus-Rebalancing}.
\newblock
\newblock
\urldef\tempurl%
\url{https://doi.org/10.48550/ARXIV.2208.06046}
\showDOI{\tempurl}


\bibitem[MountainFarmer et~al\mbox{.}(2022)]%
        {mountainfarmer22joe}
\bibfield{author}{\bibinfo{person}{MountainFarmer}, \bibinfo{person}{Louis},
  \bibinfo{person}{Hanzo}, \bibinfo{person}{Wawa}, \bibinfo{person}{Murloc},
  {and} \bibinfo{person}{Fish}.} \bibinfo{year}{2022}\natexlab{}.
\newblock \bibinfo{booktitle}{\emph{JOE v2.1 Liquidity Book}}.
\newblock
\urldef\tempurl%
\url{https://github.com/traderjoe-xyz/LB-Whitepaper/blob/main/Joe%20v2%20Liquidity%20Book%20Whitepaper.pdf}
\showURL{%
Retrieved Sept 18, 2023 from \tempurl}


\bibitem[Nezlobin(2023)]%
        {Nezlobin2023}
\bibfield{author}{\bibinfo{person}{Alex Nezlobin}.}
  \bibinfo{year}{2023}\natexlab{}.
\newblock \bibinfo{booktitle}{\emph{Twitter thread}}.
\newblock
\urldef\tempurl%
\url{https://twitter.com/0x94305/status/1674857993740111872}
\showURL{%
Retrieved Dec 3, 2023 from \tempurl}


\bibitem[Othman et~al\mbox{.}(2013)]%
        {othman2013practical}
\bibfield{author}{\bibinfo{person}{Abraham Othman}, \bibinfo{person}{David~M
  Pennock}, \bibinfo{person}{Daniel~M Reeves}, {and} \bibinfo{person}{Tuomas
  Sandholm}.} \bibinfo{year}{2013}\natexlab{}.
\newblock \showarticletitle{A practical liquidity-sensitive automated market
  maker}.
\newblock \bibinfo{journal}{\emph{ACM Transactions on Economics and Computation
  (TEAC)}} \bibinfo{volume}{1}, \bibinfo{number}{3} (\bibinfo{year}{2013}),
  \bibinfo{pages}{1--25}.
\newblock


\bibitem[Posner and Weyl(2018)]%
        {posner2018radical}
\bibfield{author}{\bibinfo{person}{Eric~A. Posner} {and}
  \bibinfo{person}{E.~Glen Weyl}.} \bibinfo{year}{2018}\natexlab{}.
\newblock \bibinfo{booktitle}{\emph{Radical Markets: Uprooting Capitalism and
  Democracy for a Just Society}}.
\newblock \bibinfo{publisher}{Princeton University Press},
  \bibinfo{address}{Princeton, NJ}.
\newblock


\bibitem[Rao and Shah(2023)]%
        {trianglefees2023}
\bibfield{author}{\bibinfo{person}{Rithvik Rao} {and} \bibinfo{person}{Nihar
  Shah}.} \bibinfo{year}{2023}\natexlab{}.
\newblock \bibinfo{title}{Triangle Fees}.
\newblock
\newblock
\showeprint[arxiv]{2306.17316}~[q-fin.MF]


\bibitem[Zhou et~al\mbox{.}(2021)]%
        {zhou2021high}
\bibfield{author}{\bibinfo{person}{Liyi Zhou}, \bibinfo{person}{Kaihua Qin},
  \bibinfo{person}{Christof~Ferreira Torres}, \bibinfo{person}{Duc~V Le}, {and}
  \bibinfo{person}{Arthur Gervais}.} \bibinfo{year}{2021}\natexlab{}.
\newblock \showarticletitle{High-frequency trading on decentralized on-chain
  exchanges}. In \bibinfo{booktitle}{\emph{2021 IEEE Symposium on Security and
  Privacy (SP)}}. IEEE, \bibinfo{pages}{428--445}.
\newblock


\end{thebibliography}
}

\appendix

\section{Proofs}

\begin{proof}[\bf\sffamily Proof of \Cref{lem:fixed-fee-eq}]
  Given fixed $f \in [0,f_{\max}]$, define $G(L) \defeq  f H_0(f,L) - \AP_0(f) - r$. By
  \Cref{as:noise-trader}(\ref{pt:nt-2}) and \Cref{as:arb-profits}, this is a continuous
  function, and
  \[
    \lim_{L\tends 0} G(L) = \infty,\quad\lim_{L\tends \infty} G(L) = -\AP_0(f) - r < 0.
  \]
  By the intermediate value theorem, there exists $L^*>0$ with $G(L^*)=0$, comparing with
  \eqref{eq:pnlff}, this must be an equilibrium. This equilibrium is unique since $G(\cdot)$ is
  strictly monotonic.
\end{proof}

\begin{proof}[\bf\sffamily Proof of \Cref{th:liq-dom}]
  First, we will prove that the equilibrium  $(R^*,L^*)$ exists. Solving for $R^*$ in the
  condition $\PNLMGR(R^*,L^*)=0$ and substituting into  $\PNLLP(R^*,L^*)=0$, we have that $L^*$
  must satisfy $G(L^*)=0$ where
  \[
    G(L) \defeq
    \max_{f\in[0,f_{\max}]}\
    f H_0(f,L) - \AE_0(f) - r.
  \]
  From \Cref{as:noise-trader} and \Cref{as:arb-profits}, this function is continuous, and
  \[
    \lim_{L\tends \infty} G(L)
    =
    \max_{f\in[0,f_{\max}]}\
    - \AE_0(f) - r
    =
    - \AE_0(0) - r < 0.
  \]
  \[
    \lim_{L\tends 0} G(L)
    \geq \lim_{L\tends 0} f  H_0(f,L) - \AE_0(f) - r
    = \infty > 0,
  \]
  for any $f \in (0,f_{\max}]$. By the intermediate value theorem, an equilibrium $(R^*,L^*)$ must
  exist.

  In order to compare with the ff-AMM, define
  \[
    L_{\max} \defeq \max_{f\in[0,f_{\max}]}\ \LFF(f),
  \]
  to be the maximum level of liquidity that can be achieved in the ff-AMM, and denote by $f^*$ the
  maximizing fee.
  Define
  \[
    R_{\max}  \defeq (\AP_0(0) + r) V(L_{\max}).
  \]
  From the zero profit condition  $\PNLLP(R_{\max},L_{\max})=0$, it is clear that a rent
  payment of $R_{\max}$ will incentivize liquidity $L_{\max}$.
  Under such a
  rent payment, the pool manager earns profits
  \[
    \begin{split}
      \PNLMGR(R_{\max},L_{\max})
      & =
        \left(\AP_0(0) +  \max_{f\in[0,f_{\max}]}\   f H_0(f,L_{\max}) - \AE_0(f) \right) V(L_{\max})
      - R_{\max}
      \\
      & \geq
        \left(\AP_0(0) +  f^* H_0(f^*,L_{\max}) - \AE_0(f^*) \right) V(L_{\max})
        - R_{\max}
      \\
      & =
        \left(\AP_0(0) +   \AP_0(f^*) + r - \AE_0(f^*) \right) V(L_{\max})
        - R_{\max}
      \\
      & =
        \left(\AP_0(0) +   \AP_0(f^*) + r - \AE_0(f^*) \right) V(L_{\max})
        - (\AP_0(0) + r)  V(L_{\max}) \\
      & = \left(\AP_0(f^*) - \AE_0(f^*) \right) V(L_{\max})
      \\
      & > 0.
    \end{split}
  \]
  Here, the first inequality follows from the suboptimality of $f^*$ for the am-AMM, and we also
  use the fact that $\PNLFF(f^*,L_{\max}) = 0$ since it is an equilibrium for the ff-AMM. The last
  equality follows from \Cref{as:arb-excess}.\todo{need to argue that $f^* > 0$}

  Since the pool manager earns positive profits when the rent is $R_{\max}$, and equilibrium rent
  $R^*$ musty satisfy $R^* > R_{\max}$, therefore and equilibrium fee-auction AMM utility
  $L^* > L_{\max}$.
\end{proof}

\section{Withdrawal Fees} \label{appendix:withdrawal}

Here, we show that a withdrawal fee of $1 - \frac{2 \cdot \sqrt{f_{cap}}}{f_{cap}+1}$ is sufficient to protect managers from strategic liquidity withdrawals. With a fee cap of 1\%, this comes out to approximately 0.00001238\%, or about 0.1238 basis points — a \$12.38 fee on a \$1 million position.

Our goal is to set a withdrawal fee to prevent opportunistic withdrawals of liquidity (in response to arbitrage opportunities) from reducing manager profits. Under our assumptions, the manager's profit from a given arbitrage opportunity is capped once the price moves by $f_{cap}$, the maximum allowable swap fee. Any larger price move results in MEV for the block proposer (since any arbitrageur can capture the excess portion with a swap), rather than profit for the manager. Therefore, we only need to set a withdrawal fee that is greater than the value from arbing that liquidity in response to an increase of a factor of $f_{cap}$. (A decrease by a factor of $f_{cap}$ will result in a smaller arbitrage opportunity.)

For this proof, we use the same conventions as \Cref{theory}, where reserves are expressed in terms of two assets $x$ and $y$, asset $y$ is defined as the num\'eraire for all prices and valuations, and ``liquidity'' for a position is defined as $\sqrt{xy}$ where $x$ and $y$ are the position's reserves of assets $x$ and $y$, respectively.

Suppose without loss of generality that a liquidity provider is providing 1 unit of liquidity, and the current midpoint price on the AMM is $p_{amm}$. Since $p_{amm} = \frac{y_{now}}{x_{now}}$ and {$x_{now} \cdot y_{now} = 1$, the liquidity provider's current reserves of assets X and Y are $x_{now} = \frac{1}{\sqrt{p_{amm}}}$ and $y_{now} = \sqrt{p_{amm}}$.

We suppose price has increased by $f_{cap}$, making the true price $p_{true} = f_{cap} \cdot p_{amm}$.

The current valuation of the liquidity is:

\begin{equation}
    v_{now} = p_{true} \cdot x_{now} + y_{now} = f_{cap} \cdot p_{amm} \cdot \frac{1}{\sqrt{p_{amm}}} + \sqrt{p_{amm}} = (1 + f_{cap}) \cdot \sqrt{p_{amm}}
\end{equation}

The valuation of the liquidity after the manager arbitrages the pool to $p_{true}$ ($f_{cap} \cdot p_{amm}$) would be:

\begin{equation}
    v_{after} = f_{cap} \cdot p_{amm} \cdot \frac{1}{\sqrt{f_{cap} \cdot p_{amm}}} + \sqrt{f_{cap} \cdot p_{amm}} = 2 \cdot \sqrt{f_{cap} \cdot p_{amm}}
\end{equation}

To prevent strategic withdrawal from being profitable for liquidity providers at the expense of managers, we can impose a withdrawal fee of $\frac{v_{now} - v_{after}}{v_{now}}$, which simplifies to:

\begin{equation}
  f_{withdrawal} = \frac{v_{now} - v_{after}}{v_{now}} = 1 - \frac{2 \cdot \sqrt{f_{cap}}}{1 + f_{cap}}
\end{equation}

%%% Local Variables:
%%% mode: latex
%%% TeX-master: "am-amm"
%%% End:

\end{document}